\documentstyle[prl,aps,multicol]{revtex}
\input{epsf}
\begin{document}
\draft
\title{Quantum Fractal Fluctuations}  
\author{Giuliano Benenti$^{(a,b)}$, Giulio Casati$^{(a,b,c)}$, 
Italo Guarneri$^{(a,b,d)}$, and Marcello Terraneo$^{(a,b)}$}  
\address{$^{(a)}$International Center for the Study of Dynamical 
Systems,} 
\address{
Universit\`a degli Studi dell'Insubria, Via Valleggio 11,  
22100 Como, Italy} 
\address{$^{(b)}$Istituto Nazionale per la Fisica della Materia, 
Unit\`a di Milano, Via Celoria 16, 20133 Milano, Italy}   
\address{$^{(c)}$Istituto Nazionale di Fisica Nucleare, 
Sezione di Milano, Via Celoria 16, 20133 Milano, Italy}   
\address{$^{(d)}$Istituto Nazionale di Fisica Nucleare, 
Sezione di Pavia, Via Bassi 6, 27100 Pavia, Italy}   
\date{\today}
\maketitle

\begin{abstract}
We numerically analyse quantum survival probability fluctuations 
in an open, classically chaotic system. In a quasi-classical regime, and in
the presence of classical mixed phase space, such fluctuations are 
believed to exhibit a fractal pattern, on the grounds of semiclassical 
arguments.
In contrast, we work in a classical regime of complete chaoticity, 
and in a deep quantum regime of strong localization. We provide 
evidence that fluctuations are still fractal,  due to 
the slow, purely quantum algebraic  decay in time produced by 
dynamical localization.
 Such findings considerably enlarge  the scope of the existing theory.

\end{abstract}

\begin{multicols}{2}
\narrowtext

The study of mesoscopic conductance fluctuations has gained important 
insights from the analysis of the underlying classical dynamics
\cite{jal}.
In particular, it has been found that some statistical properties 
of quantum fluctuations are semiclassically 
determined by the classical law which rules the decay in time of the 
survival probability inside the open 
chaotic system\cite{smila}. Building on this general relation, it was eventually 
predicted that, if the classical decay law is algebraic, due to 
the presence of residual stable islands in the classical phase space, 
then the quantum fluctuation pattern should display a fractal structure 
with a fractal dimension directly related to the classical decay exponent
\cite{ketz}. 
Such fractal structure of quantum 
fluctuations has been interpreted as a quantal manifestation of
classical fractality  associated  with hierarchical 
structures of surviving stable islands.
This theoretical prediction has received experimental support\cite{expe}, 
and has 
been numerically confirmed on a dissipative model system, 
where coupling to continuum is simulated by complete absorption of that part of the
wave packets which propagates outside the interaction region\cite{casa,spain}.

In such a theoretical frame, quasi-classicality of the quantum dynamics 
appears essential in order that slow classical decay be quantally 
reproduced over sufficiently long time scales. In contrast, in a deep quantum 
regime, localization effects would become dominant, rapidly effacing any 
memory of the underlying classical dynamics.

In this Letter we provide evidence that  fractality of quantum fluctuations 
nevertheless survives even in such strongly  non-classical 
situations. In the model systems presently to be described, the classical 
phase space has no 
significant stable islands, the classical motion is diffusive, and  
 the classical decay is exponential; therefore, in the quantum weakly 
localized regime, one observes non-fractal, Lorentz-correlated 
conductance fluctuations\cite{borgo}. In contrast, we work in a 
strongly 
localized regime, and observe fractal fluctuations, apparently unrelated to 
any classical phase-space structure. Decisive in this respect is the fact 
that the {\it quantum} decay, albeit highly non-classical, is algebraic 
 due to purely quantum localization effects, as 
recently predicted\cite{dima,hatom}. As a matter of fact, we find that 
 the fractal dimension of quantum fluctuations is connected to the algebraic
decay exponent in exactly the same way predicted by the semiclassical theory
\cite{ketz,casa}.
 It therefore appears that the quantum 
fluctuation pattern bears the same relation to the {\it quantum} 
probability decay, which was originally predicted by semiclassical arguments,
even though such decay is not by any means quasi-classical.

We consider the kicked rotator model \cite{fishman} with absorbing boundary 
conditions. Classical dynamics is described by the  standard map:
\begin{equation}
\label{clrot}
\bar{I}  =  I + k \sin \theta, \quad 
\bar{\theta}  =  \theta + T \bar{I}, 
\end{equation} 
with absorption for $\bar{I} < 0 $ and for $\bar{I} > N$.
Classical dynamics only depends on the scaling parameter $K=kT$.
The corresponding quantum map is obtained by the substitution $ I \to
\hbar \hat{n} = -i \hbar \partial / \partial \theta$. 
 We denote $\psi_t$ the state vector immediately before 
the $t-$th kick, $T$ the kicking period, and we set $\hbar=1$.
Throughout this Letter, we will treat the kick counter
$t$ as a discrete time variable in units of $T$. 
The discrete-time evolution from 
time $t$ to time $t+1$ is described in the classical case by 
(\ref{clrot}) and in the quantum case by :
\begin{equation} 
\label{rot}
{\psi_{t+1}}=\hat{P}\tilde\psi_{t+1}=\hat{P}e^{-iT(\hat{n}+\phi)^2/2} 
e^{-ik\cos\hat{\theta}}\psi_t.  
\end{equation} 
The state   
   $\tilde\psi_{t+1}$ is obtained from $\psi_{t}$ by means of the unitary 
one-period propagator of the kicked rotor. The operator  
 $\hat{P}$ then projects  $\tilde\psi_{t+1}$  over states in a fixed 
interval of $n$ values ($0 \le n \le N$), so it describes complete
deletion (absorption) of the  
part of the wave packet which propagates outside the given interval. 
 The parameter $\phi$ can be interpreted as an  
Aharonov-Bohm flux
through the ring parametrized by the coordinate $\theta$.

We consider the case $K=kT=7$,  $k=5$, with 
absorption for $n< 0$ and $n>N=250$. With such parameter values, 
the classical phase space has no  
significant island of stability \cite{note}. We consider a statistical 
ensemble of orbits starting at $t=0$ with $I=0$ and randomly 
distributed phases $\theta$. The survival probability 
$P(t)$ is the fraction of the ensemble which hasn't been absorbed 
at time $t$; equivalently, $P(t)$ is the integrated distribution of 
exit times of orbits in the ensemble.

\begin{figure}
\centerline{\epsfxsize=7.2cm\epsffile{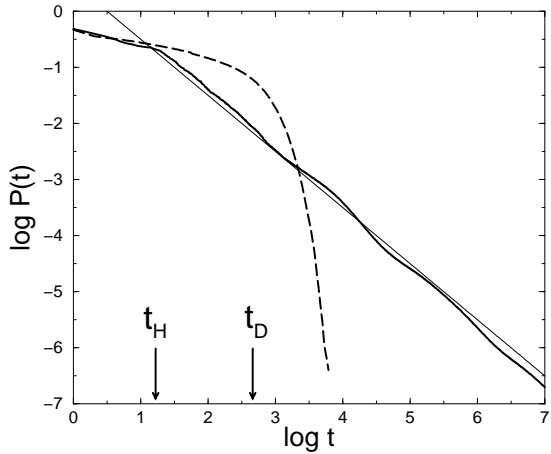}}
\caption{Classical (dashed line) and quantum (thick solid line) survival 
probability for the kicked rotator model with $K=7$, $k=5$, $\phi=0$. 
The evolution starts at $n_0=0$ and the probability is absorbed for 
$n< 0$ and $n>250$. The thin solid line has slope one.  
Here and in the following figures the logarithms are decimal. 
} 
\label{fig1} 
\end{figure}
 
\vskip -0.5cm 
\begin{figure}
\centerline{\epsfxsize=7.5cm\epsffile{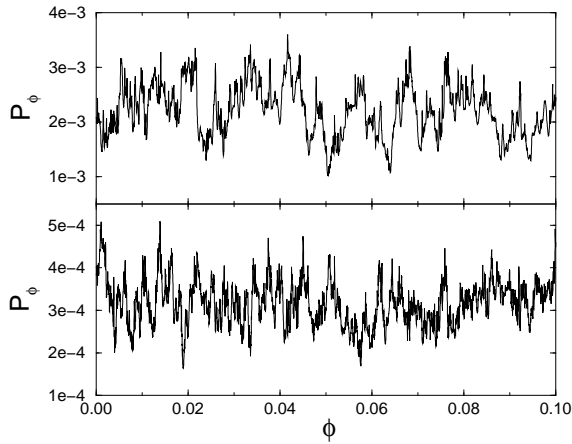}}
\caption{Survival probability vs flux $\phi$ at fixed time 
$t=10^3$ (above) and $t=10^4$ (below), with parameter values as 
in Fig.\ref{fig1}. 
} 
\label{fig2} 
\end{figure}

The classical decay of $P(t)$  is well described 
by a diffusion process, practically insensitive to the chosen 
value of $\phi$. As 
shown in Fig.\ref{fig1}, after a diffusion time 
$t_D\approx N^2/D\propto N^2/k^2$ ($D$ diffusion rate), 
the survival probability exponentially decays  with time:
$P(t)\propto\exp(-\gamma t)$, with 
$\gamma\approx 1/t_D\approx 2.2\times 10^{-3}$. In contrast, quantum dynamics 
is strongly localized, as the localization length $\ell\approx k^2$  
is much less than the ``sample size'' $N$. Such strong localization 
causes a slow decay of the quantum survival probability $P_{\phi}(t)$ 
  The latter is  
defined as the total probability on states $0<n<250$ at time $t$, 
given that at $t=0$ the state was $n_0=0$. After the ``Heisenberg time'' 
 $t_H
$, it 
decays proportional to $1/t$ up to a time $t_{max}\sim\exp(2N/{\ell})$ 
\cite{dima}, as shown in 
Fig.\ref{fig1} (in this case, $t_{max}\sim 5\times 10^8$). 

Though the exponent $1$ of algebraic quantum decay is independent 
of the value of $\phi$, the quantum survival probability 
$P_{\phi}(t)$ at a fixed 
time $t$ is sensitively dependent on $\phi$. The fluctuation pattern 
exhibited by 
the graph of 
$P_{\phi}(t)$ versus $\phi$ at fixed $t$ (two examples are given 
in Fig.\ref{fig2})  
is precisely the object of our analysis, at values of $t$ up to 
$10^4$, and for $10^4$ values of $\phi$ in the interval 
$[0,0.1]$. 
First we have computed autocorrelation functions:
\begin{equation} 
\label{corr0}
C(\delta\phi)=<P_\phi(t)P_{\phi+\delta\phi}(t)>_{\phi}.  
\end{equation}  
Such correlations play a key role in the existing semiclassical theory of
fractal fluctuations, which  consists of three steps:

\noindent
(i)  correlation functions are related 
via Fourier transform 
to the classical decay of the canonical variable
conjugate to $\phi$\cite{smila,casa}, notably, to the 
function
$P_{cl}(\Theta)$ yielding the probability distribution of the angle $\Theta$ accumulated 
by the rotor at the time of exit from the finite sample (``accumulated'' 
means multiples of $2\pi$ included). In particular, if 
 $P_{cl}(\Theta)$ decays as $\Theta^{-\alpha}$ at large 
$\Theta$, then 
\begin{equation} 
\label{corr}
C(\delta\phi)\sim C(0)-\hbox{const}|\delta\phi|^\alpha 
\end{equation} 
at small $\delta\phi$.

\noindent
(ii) It is physically intuitive, and 
numerically confirmed \cite{casa}, that the decay exponent $\alpha$  
coincides with the decay exponent of the classical survival probability. 

\noindent
(iii) a signal-theoretic argument \cite{mandelbrot} 
allows to conclude from (i) and (ii)
that the 
fluctuation graph has a fractal dimension given by
\begin{equation} 
\label{frac}
D=2-\frac{\alpha}{2} \quad (\hbox{for }\alpha\leq 2),   
\end{equation} 
where $\alpha$ is the time decay exponent of the classical survival 
probability.
Step (iii) actually invokes certain statistical properties of 
fluctuations, which were assumed to be ensured by the chaotic nature 
of classical dynamics\cite{ketz1}.

\begin{figure}
\centerline{\epsfxsize=7.2cm\epsffile{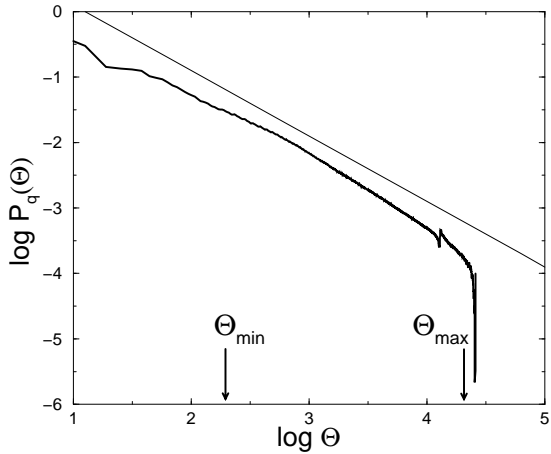}}
\caption{Quantum integrated distribution of the accumulated angle $\Theta$ 
at the exit time, computed over a total time $t=600$, parameter 
values as in Fig.\ref{fig1}.  The straight line has slope one.  
} 
\label{fig3} 
\end{figure}

In our case, semiclassical arguments are invalid,  
because of strong quantum localization. In addition, there is no 
classical decay exponent $\alpha$, because classical 
decay is exponential. So the above theory doesn't apply; still,   
numerical results to be described below demonstrate that (i), (ii)
and (iii) are 
still legitimate in our 
highly non semiclassical case, too, provided one replaces the classical 
distribution functions $P(t)$,$P_{cl}(\Theta)$ by the corresponding 
quantum distributions $P_{\phi}(t)$,$P_q(\Theta)$, and 
$\alpha$ by the exponent of the {\it quantum} algebraic decay. 
In order to compute $P_q(\Theta)$ we perform a lifting of our model
(\ref{rot}), that is, we translate the 
 model (\ref{rot}) of a kicked particle on a circle into the model of a
kicked particle on the line, described by the coordinate $\Theta$. To simulate
the corresponding dynamics we restrict $\Theta$ inside a large, yet finite 
box, with periodic boundary conditions.  
At every 
evolution step $t$ we compute the function $|\psi_t(\Theta)-
\tilde\psi_t(\Theta)|^2$, which yields the  probability loss occurring at 
time $t$ and position $\Theta$ due to absorption. 
Summing this quantity over $t$ gives the 
total probability lost at position $\Theta$, that is, the probability 
distribution of the  angle accumulated at the exit time. The corresponding 
integrated distribution is  shown 
in Fig.\ref{fig3} and displays a decay $\propto \Theta^{-1}$. 
The faster decay 
in the rightmost part of Fig.\ref{fig3} is due to the finite total time 
and to the finite $\Theta-$box used in the integration. Thus, at any given 
time $t$, the distribution of accumulated exit angles up to that time 
can be assumed to decay proportional to $\Theta^{-1}$ in a range 
$\Theta_{min}<\Theta<\Theta_{max,t}$, with $\Theta_{max,t}$ increasing with 
$t$ as long as $t<t_{max}$. 
One can estimate $\Theta_{min}$ as the angle accumulated until  
the time $t_H\approx\ell$. 
 This gives     
$\Theta_{min}\sim \int_0^{t_H T} \sqrt{(D/T) \tau} d \tau 
\sim (2/3)t_H^{2}T$. 
Assuming a maximal momentum 
$\sim\ell$ for the rotor started at $n_0=0$, 
a rough estimate for $\Theta_{max,t}$ is provided by $\ell tT$ 
($\Theta_{min}\sim 2 \times 10^2$ and $\Theta_{max,t}\approx 
2.1\times 10^4$ for the case of Fig.3).  

\begin{figure}
\centerline{\epsfxsize=7.5cm\epsffile{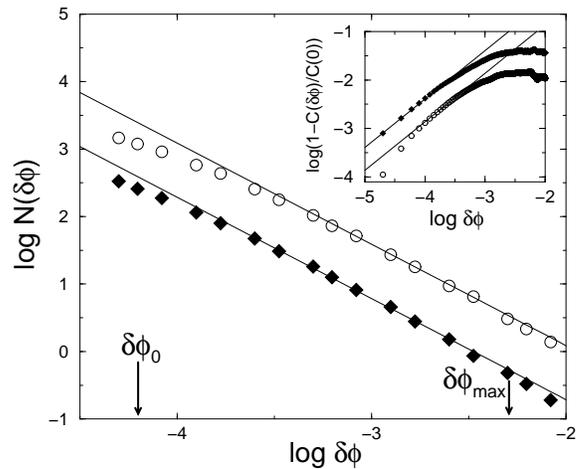}}
\caption{Fractal analysis for the kicked rotator model, with 
parameter values as in Fig.\ref{fig1}. The graph of $P_{\phi}(t)$ 
vs $\phi$ was covered by boxes of side $\delta\phi$, and the largest 
excursion of $P_\phi(t)$ in each strip was recorded; summing over all 
strips and dividing the result by $\delta\phi$, we obtained $N(\delta\phi)$. 
Data are shown for $t=10^3$ (circles) 
and $t=10^4$ (diamonds). The straight lines correspond 
to the fractal dimension $D=1.5$ expected for $\alpha=1$.  
The inset shows the correlation function $C(\delta\phi)$ at the 
same times. The straight lines correspond to 
$1-C(\delta\phi)/C(0)\propto |\delta\phi|^\alpha$, with $\alpha=1$.  
} 
\label{fig4} 
\end{figure}

Numerically computed correlations (\ref{corr0}) shown in the inset of 
Fig.\ref{fig4}, 
at two different times, indicate 
 that relation (\ref{corr}) is still valid in our 
strongly localized case\cite{borgo1}. 
We have computed the fractal dimensions   
of the corresponding graphs of $P_{\phi}(t)$ versus $\phi$ shown 
in Fig.\ref{fig2} using the box-counting algorithm described in 
\cite{expe,casa}. The results are presented in  
Fig.\ref{fig4} and demonstrate that the graph 
has a fractal dimension 
$\approx 1.5$ over a significant range of $\delta\phi$ scales, 
as predicted by equation (\ref{frac}).
This range roughly lies in between scales 
$\delta\phi_{min,t}$,
$\delta\phi_{max}$. The latter scale is inverse to 
$\Theta_{min}$. The former scale can be estimated 
by  $\Theta_{max,t}^{-1}$ (hence it decreases as $t$ increases),
as long as it remains above a minimal $\delta\phi_0\sim 
2\pi/(N^2T)$ scale. This scale is imposed 
 by the finiteness of the sample, and can be explained as follows. 
The evolution operator (\ref{rot}) has complex eigenvalues inside the unit 
circle. Varying $\phi$ causes these eigenvalues to move, and their maximal 
shift under a change $\delta\phi$ is estimated by $TN\delta\phi$. 
 Any time one moving eigenvalue  comes close to the unit circle, a local maximum  in the graph of $P_{\phi}(t)$ 
vs $\phi$ appears. The number of peaks produced by a single moving 
eigenvalue as $\phi$ changes from $0$ to $1$  is at most $\sim TN/(2\pi)$. 
As there are $N$ eigenvalues, the total number of peaks is $\sim 
TN^2/(2\pi)$, with an average $\phi-$ spacing $\sim \delta\phi_0=
(2\pi)/(N^2T)$. 
As the fractal structure of the graph is due to the superposition of many 
tiny peaks originated in this way, no fractality can  
be expected below the $\delta\phi_0$ scale.

\begin{figure}
\centerline{\epsfxsize=7.5cm\epsffile{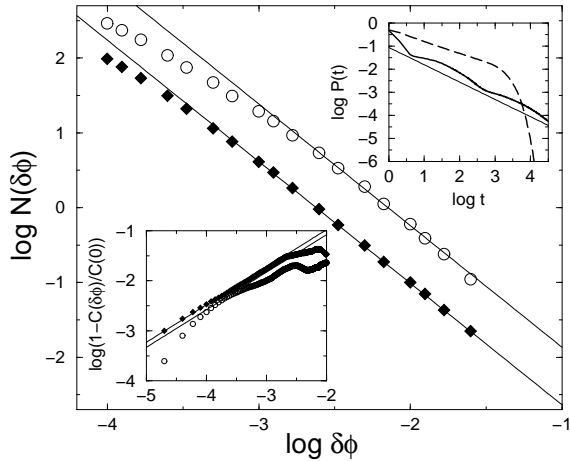}}
\caption{
Same as in Fig.\ref{fig1} (upper inset) and Fig.\ref{fig4} 
but for the discontinuous map (\ref{disc}). 
The straight lines correspond to a power law decay of the 
quantum survival probability with $\alpha=3/4$ (upper inset), 
a correlation function $C(\delta\phi)\propto |\delta\phi|^{\alpha}$ 
(lower inset) and a fractal dimension $D=2-\alpha/2=13/8$. 
The $x$-axis in the lower inset is the same as in the 
main figure. 
} 
\label{fig5} 
\end{figure}

Further confirmation for the validity of the above illustrated scenario was 
obtained 
by analyzing the following variant of the basic model (\ref{rot}): 
\begin{equation}
\label{disc}  
\psi_{t+1}=\hat{P} e^{-iT(\hat{n}+\phi)^2/2} 
e^{-ik|\cos\hat{\theta}|}\psi_t.  
\end{equation} 
which differs from (\ref{rot}) because the kicking potential has now a
discontinuous derivative. Localization is in this case algebraic, 
as eigenfunctions decay like $1/n^2$. The argument used to predict 
the $t^{-1}$ decay of the survival probability in the case of 
exponential localization \cite{dima}
yields an
asymptotic decay $\propto t^{-3/4}$. The numerical analysis 
illustrated above for the case of exponential localization was replicated 
for this model, too, yielding the results shown  in Fig.\ref{fig5}.
Fractal analysis of the fluctuation graph yields good agreement 
over a broad range with the theoretical $D$  
value of $13/8$ predicted using equation (\ref{frac}) and 
the decay exponent $\alpha=3/4$.

In conclusion, we have provided numerical evidence for fractal fluctuations 
of the quantum survival probability in a classically chaotic system 
in the regime of strong localization, and in the absence of significant 
classical critical structures.  
Whereas the original theory \cite{ketz} predicted fractal {\it conductance} 
fluctuations, our present results are about fluctuations 
of a different quantity, namely  {\it survival probability}. 
As both quantities ultimately  reflect the fluctuations  of the 
scattering matrix, their fluctuations should be similar in nature. As a 
matter of fact, the very same semiclassical arguments used to predict  
conductance  fluctuations could be adapted to the survival 
probability, too \cite{casa}.  The main difference is that a scan of  $P_{\phi}(t)$ 
at finite however large $t$ cannot achieve the $\phi-$resolution 
exhibited by conductance at fixed (quasi-)energy. On account of 
the  effect of increasing $t$ reported earlier in this Letter, such  
coarsening is unlikely to  ``fractalize'' otherwise non-fractal 
fluctuation patterns. 
In any case,  
survival probability is a meaningful quantity in a broader class 
of problems than electronic transport in semiconductor structures. 
 The scenario we have analyzed in this Letter 
can also be exported to  other realistic problems, where 'conductance' 
is instead  a problematic concept. As an example we quote  
microwave ionization of Rydberg atoms \cite{hatom}. This may open a way for 
experimental observation of localization-induced fractal fluctuations in 
atomic physics. 

Our results show that quantum fluctuations may be fractal even in 
situations where the existing semiclassical theory does not apply.
This does not command a reinterpretation of existing experimental data, 
which were obtained in situations far from the strongly localized regime 
considered in this Letter. Nevertheless it signals that the 
current understanding of the deep quantum mechanisms 
responsible for fractal fluctuations is still far from complete.

Support from the Progetto Avanzato INFM ``Quantum transport 
and classical chaos'' is gratefully acknowledged.

\end{multicols}


\begin{references}
\bibitem{note} Even though accelerator modes are present at $K=7$, 
they do not significantly affect the transport properties discussed in 
this Letter.
\end{references}
\end{document}